\documentclass{article}
\title{Exact Coupling Of Event Horizons In Curved Spacetime Heterostructures. Application To Black-Hole Physics.}
\author{Fredrick Michael *}
\begin{document}
\maketitle

\abstract{Recently we have discussed the generalized parametrized Klein-Gordon equation for curved spacetime. We have also discussed its derivation from several approaches, the direct Feynman parametrization, the state function entropy or equivalently the information theory approach, and the stochastic differential equation approach. We have even suggested a generalization of the statistics of the entropy to the generalized entropies and derived the particular nonextensive statistics parametrized Klein-Gordon equation, and discussed its nonlinear FPE replacement of the complicated Gibbs-Boltzmann statistics entropy derived analog with complicated nonlinear potential or drift and diffusion coefficients. In this article we apply these previously derived results to the quantum transport in abruptly coupled curved space-time heterostructures, applied here specifically to Black-Hole event horizon coupling to normal curved space-time. We derive the coupling self energy, and the Garcia-Molliner surface Green's functions from which we can calculate the surface area and entropy. We then derive the nonequilibrium transport equations for the radiation from the Black-Hole. We discuss the theory application to Worm Holes and quantum analogues. } 
    
\section{Introduction}

\section{derivation}
The parametrized Klein-Gordon equation can be obtained from the maximum entropy equation as the maximization of the entropy $<S>=-\int PlnP$ with the moments $<\vec x^2>=\int \vec x^2 P(\vec x, \lambda)d\vec x$ multiplied by the Lagrange multipliers as \begin{equation}
\delta[<S(\lambda)>]+\delta[\sum\limits_{i=1}^{4}\beta_i ({x_i} ^2)]=0
\end{equation}
The maximization allows us to derive the least biased probability distribution function
\begin{equation}
P(\vec x,\lambda)=N(\lambda){e^{-\sum\limits_{i=1}^{4}\beta_i ({x_i} )^2 }} \label{eq2}
\end{equation}
The Lagrange multipliers are evaluated by the identity  
$-\frac{1}{Z}\frac{\partial Z}{\partial \beta_i}=<(x_i)^2>=2 D_i \lambda$ . The partition function is the inverse of the normalization $Z=\frac{1}{N}$.
The PDE of the evolution of the least biased distribution Eq.(\ref{eq2}) is for the parametrization $-\beta \le \lambda \le \beta$
\begin{equation}
\frac{\partial P}{\partial \lambda}=\frac{1}{2}\sum\limits_{i=1}^{4}\frac{\partial {\beta_{o_i}} P}{\partial {x_i} ^2}.
\end{equation}
The Lagrange multipliers which are the inverses of the diffusion coefficients are when set to the coordinate $D_{i=1,2,3}=2 $, $D_4 = \frac{2}{c^2}$ and upon taking the Laplace transform (or Fourier transform with $\lambda-> -i \lambda $ with appropriate analytic continuation) $ P(\vec x, \lambda)=\int P(\vec x, s) e^{- \lambda s} ds$ becomes with $s=\frac{m^2 c^4}{\hbar ^2}$ the massive Klein-Gordon equation
\begin{equation}
\frac{\partial P}{\partial {x} ^2}+\frac{\partial P}{\partial {y} ^2}+\frac{\partial P}{\partial {z} ^2}+ \frac{1}{c^2}\frac{\partial P}{\partial {(-it)} ^2}  =\frac{m^2 c^4}{\hbar^2} P
\end{equation}
The generalization to the curved space time is done by the inclusion of diffusion coefficients from  the metric tensor. The diffusion coefficients are the inverses of the diagonal metric tensor. Alternatively the curved spacetime potential is utilized, the Shroedinger-like PDE is obtained, a transformation to the Fokker-Planck PDE is made and therefore drift and diffusion coefficients are obtained. Utilizing the Girsanov theorem transforms that in turn to a drift free form which has the coordinates transformed as the drift dependent coordinates Alternatively SDE transformations can be made.

The diffusion coefficients we will be working with are the diffusion coefficients from the inverses of metric tensor of the Black-Hole curved space-time.

The Schwarzchild metric in Cartesian coordinates is 
\begin{equation}
d \tau^2 = - \frac{(1-\frac{M  G}{c^2 r})^2} {(1+\frac{M  G}{c^2 r})^2} dt^2 + {(1+\frac{M  G}{c^2 r})^2} (dx^2 + dy^2 + dz^2)
\end{equation}

The metric space-like differential equation is , with $t^* = -it$
\begin{equation}
d\vec \tau= {A}_x (\vec x)dx \:\: \hat i +  { A}_y (\vec x)dy \:\: \hat  j +  { A}_z (\vec x)dz \:\: \hat k +  
{ A}_t (\vec x)d{t^*} \:\: {\hat t}
\end{equation}
with the stochastic differential equations
\begin{eqnarray}
d\vec \tau=d {\vec W}(\lambda) \;\;\;\;\;\;\;\;\;\;\;\;\;\;\; \\ \nonumber
dx=\sqrt{{(1+\frac{M  G}{c^2 r})^{-2}} }   \:\: d {W_x}(\lambda) \\ \nonumber
dy=\sqrt{{(1+\frac{M  G}{c^2 r})^{-2}} }   \:\: d {W_y}(\lambda) \\ \nonumber
dz=\sqrt{{(1+\frac{M  G}{c^2 r})^{-2}} }    \:\:d {W_z}(\lambda) \\ \nonumber
d{t^*}= \frac{(1+\frac{M  G}{c^2 r})} {(1-\frac{M  G}{c^2 r})}    \:\: d {W_t}(\lambda) \\ \nonumber
\end{eqnarray}
This equation when the diffusion coefficients are multiplied to the left and the metric product taken becomes the space-like metric, and the delta correlated noise terms $(dW(\lambda))^2=d\lambda$  of the space-like metric vector $d\vec \tau $. The delta correlated noise terms $ (d {W_t}(\lambda))^2 + ( d {W_x}(\lambda))^2 +( d {W_y}(\lambda))^2 + (d {W_z}(\lambda))^2 = (d {W}(\lambda))^2=d\lambda    $. That is the sum of delta correlated Wiener processes is a Wiener process, and the square of the delta correlated process is replaced by its average $d\lambda$ \cite{gardiner1}.

The SDEs are utilized to obtain the Fokker-Planck equation \cite{gardiner1} for the process, which is
\begin{equation}
\frac{\partial P(\vec x,\lambda)}{\partial \lambda}=\frac{D}{2}\sum\limits_{i=1}^{3} \frac{\partial^2 }{\partial {x_i} ^2}({{(1+\frac{M  G}{c^2 r})^{-2}} } )P+\frac{D_t}{2} \frac{\partial^2}{\partial {(-it)} ^2} (\frac{(1+\frac{M G}{c^2 r})^2} {(1-\frac{M  G}{c^2 r})^2}) P 
\end{equation}
Therefore the diffusion coefficients, with the $D=2, D_t=\frac{2}{c^2}$ factored out from the noise terms and written explicitly. The constant diffusion coefficients are now generalized to the inverses of the Schwarzchild metric. The Fokker Planck equation is solved by several methods, and we write the short time transition probability solution \cite{risken1}

    The solution of the time here evolution parameter $\lambda$ dependent Fokker-Planck equation is obtained as
\begin{equation}
P(\vec x,t; \vec x', t')=\frac{e^{-{(\vec x - \vec x' )[\sigma^{-1} ( \vec x', t')](\vec x - \vec x')}}}{\sqrt{4\pi \sigma (\vec x', t')}}  \label{eqn535}
\end{equation}

The FPE equation is also to be transformed by SDE transformations and compared to the solution Eq.({\ref{eqn535}}). We write the SDE for example the $X$ coordinate as

\begin{eqnarray}
dX=\frac{1}{2} \frac{\partial^2 X}{\partial x^2} {(1-\frac{M G}{c^2 r})^{-2}} d\lambda +  \frac{\partial X}{\partial x}\sqrt{(1-\frac{M G}{c^2 r})^{-2}} d W_x (\lambda)
\end{eqnarray}
We set the diffusion coefficient of this transformed SDE to a constant and obtain after a change of variables $x'=r'cos(\theta')sin(\phi')$ the differential 
\begin{eqnarray}
X= D'' \int\limits_{0}^{x} \sqrt {(1-\frac{M G}{{c^2} r'})^2} dx'= \;\;\;\;\;\;\;\;\;\;\;\;\;\;\;\;\;\;\;\;\;\;\;\;\;\;\;\;\;\;\;\;\;\;\;\;\;\;\;\;\;\;\;\;\;\;\;\;\;\;\;\;\;\;\;\;\;\;\;\;\;\;\;\;\;\;\;\; \\ \nonumber
{ D'' c^2} (\frac{1}{r' cos(\theta ') sin(\phi ')}) 
\frac{\partial}{ \partial \theta '}(\frac{-1}{r' sin(\phi ') }) \frac{  \partial }{\partial \phi '}  \int\int \int\limits_{0}^{x}  \sqrt {(1-\frac{M G}{c^2 r'})}   {r'^2} dr' sin(\theta ') {d\theta '} {d\phi '} \\ \nonumber
=[....] {D''  c^2} \sum_{l=0}^{inf} \int\limits_{0}^{r} \frac{{(-1)^n} (2n)! }{(1-2n) {(n!)^2} (4n) {(M G)^n}}{r'^n} \;\;\;\;\;\;\;\;\;\;\;\;\;\;\;\;\;\;\;\;\;\;\;\;\;\;\;\;\;\;\;\;\;\; \\ \nonumber
=[....]  {D''  c^2} \sum_{l=0}^{inf} \int\limits_{0}^{r} \frac{{(-1)^n} (2n)! }{(1-2n){(n!)^2} (4n) (n+1){(M G)^n}}{r^{n+1}} \;\;\;\;\;\;\;\;\;\;\;\;\;\;\;\;\;\;\;\;\;\;\;\;\;\;\;\;\;\;\;\;\;\;
\end{eqnarray}

The SDE can be written as
\begin{equation}
dX=\frac{1}{2} \frac{\partial (1+\frac{M  G}{c^2 r})}{\partial x} {(1-\frac{M G}{c^2 r})^{-2}} d\lambda + \sqrt{2D''} d W_x (\lambda)
\end{equation}
and the transformation of coordinates for $dy,dz,d\tau$ is made similarly to obtain the drift coefficients
\begin{eqnarray}
A_i ({\vec x,\lambda})=\frac{\partial (1+\frac{M  G}{c^2 r})}{\partial x_i} {(1-\frac{M G}{c^2 r})^{-2}} \\ \nonumber
A_{t^*} ({\vec x,\lambda})=\frac{\partial  \frac{{(1-\frac{M G}{c^2 r})}}{{(1-\frac{M G}{c^2 r})}} }{\partial {t^*} } \frac{{(1+\frac{M G}{c^2 r})^{2}}}{{(1-\frac{M G}{c^2 r})^{2}}}.
\end{eqnarray}
This stochastic process has the Fokker-Planck evolution, and we redefine $t^* = -it =>-ict$ and $\vec x -> x,y,z,t^*$
\begin{equation}
\frac{\partial P(\vec x,\lambda)}{\partial \lambda}=-\frac{\partial }{\partial \vec x}[\vec A P]+ \frac{D''}{2}\sum\limits_{i=1}^{4} \frac{\partial^2 P}{\partial {x_i} ^2} .
\end{equation}
The two point transitional probability for short times solves the Fokker-Planck equation with the transition time as the difference $\lambda-\lambda'= \Delta \lambda$
\begin{equation}
P(\vec x,{t^*}; \vec x', {t^*}')=\frac{e^{-\frac{{(\vec x - \vec x'- {\vec A} (\vec x',{t^*}')\Delta \lambda )^2 }}{{2 D'' \Delta \lambda}}}} {\sqrt{4\pi D'' \Delta \lambda}}.  \label{eqn45}
\end{equation}

\section{Derivation from potential }
the equations for the Schwarzchild metric and their solution have been discussed in terms of the derived drift and diffusion coefficients. An alternative method is to derive these equations from the state function entropic measure, with the gravitational potential due to the Black-Hole a constraint . We derive the least biased probability distribution for the potential $<V>$.

The potential is obtained from a Lagrangian $L=T-V$, with the variation obtaining the Euler-Lagrange equation $\frac{d}{d\tau}\frac{\partial L}{\partial \frac{d\vec x}{d\tau}}- \frac{\partial L}{\partial V}=0$. The differential for example $\frac{d x}{d\tau}=\frac{d x}{d t^*} \frac{d t^*}{d \tau}=  \frac{(1+\frac{M  G}{c^2 r})^2}{(1-\frac{M  G}{c^2 r})^2}  \frac{d x}{d t^*}$
and the differential $\frac{d}{d t^*}\frac{d t^*}{d\tau}=\frac{(1+\frac{M  G}{c^2 r})^2} {(1-\frac{M  G}{c^2 r})^2}\frac{d}{d t^*}$. The potential then for the example $x$ coordinate becomes $\frac{d}{d\tau} m \frac{dx}{d\tau}$. 

The kinetic energy has a mass term, and as we parametrized the massive Klein-gordon equation, we will need to rewrite the mass term as the parametrization $ \lambda$....and note that other parametrizations could have been made which after calculation could be taken as a limit approaching zero, and as the massive Klein-Gordon equation utilized in the photon propagator was parametrized and after calculation the mass term was taken as a limit approaching zero.

The kinetic energy term $T= \frac{m}{2} (\frac{dx}{d\tau})^2$   has a mass term, which for example can be written as $T = \frac{\hbar^4}{c^8}\sqrt{\int \frac{\partial^2}{\partial \lambda^2} e^{-s\lambda} d\lambda} $. The Euler-Lagrange equation has the force relationship $ {m} \frac{d}{d\tau}(\frac{dx}{d\tau})=\frac{\partial V}{\partial x}$  which is obtained from the discussed metric and from which we derive the relations $\frac{\partial}{\partial \lambda}[ \frac{(1+\frac{M  G}{c^2 r})^2} {(1-\frac{M  G}{c^2 r})^2}   (\frac{\partial}{\partial t^*}\frac{(1+\frac{M  G}{c^2 r})^2}{(1-\frac{M  G}{c^2 r})^2} {v_x}+   \nabla_x \frac{(1+\frac{M  G}{c^2 r})^2}{(1-\frac{M  G}{c^2 r})^2} {v_x} ^2 + \frac{(1+\frac{M  G}{c^2 r})^2}{(1-\frac{M  G}{c^2 r})^2} a_x  ] =F_x $ .

The very complex force components are simpler in the spherical polar coordinates due to the radial dependence of the potential. We write the potential as the radial $V= -\frac{GM}{r \sqrt{1-\frac{v^2}{c^2}}}$. We can rewrite this in Cartesian coordinates. The velocity squared  is $v^2 = {v_x} ^2 + {v_y} ^2 + {v_z} ^2 + {v_{t^*}} ^2$, and the radial coordinate is the modulus of the Cartesian two body distance $r=\sqrt{(x- x_o )^2 + (y- y_o )^2 + (z-  z_o )^2 + (t^*  -  {t^*}_o )^2}$.

The entropy is the, not to be confused with the spacelike metric $ds^2$, $<S(\lambda)>=-\int PlnP$, the potential $<V(\vec x, \lambda)>=\int V(\vec x) P(\vec x,\lambda)d\vec x$. The maximization yields 
\begin{equation}
P(\vec x, \lambda; \vec x \lambda')=\frac{e^{-\beta [(\vec x - \vec x'  ))^2 +\alpha V(\vec x',\lambda')]}}{Z(\Delta \lambda)}
\end{equation}
This can be rewritten by uncompleting the square of $(\vec x - \vec x' - {\hat\alpha}(\vec x',\lambda',\lambda) {\vec a} (\vec x', \lambda') )$,
\begin{equation}
P(\vec x, \lambda; \vec x \lambda')=\frac{e^{-\beta [(\vec x - \vec x' - {\hat\alpha}(\vec x',\lambda',\lambda) {\vec a} (\vec x', \lambda') )]}}{Z(\Delta \lambda)}
\end{equation}
We compare this to the short time transition probability solution Eq.(\ref{eqn45}) and note that the Lagrange multipliers are $\beta=\frac{1}{2D\Delta \lambda}$ , the partition function $Z=\sqrt{4\pi D \Delta \lambda}$ and the Lagrange multiplier we factored out, $\hat \alpha({\vec x', \lambda',\lambda})$ a complicated function we can obtain from the Fokker-Planck equation or by direct integration, and by the identity $-\frac{1}{Z}\frac{\partial Z}{\partial \alpha}=<V>$. Also we note that the flat spacetime operators we utilized can be modified to include curvature by the similar change in coordinates as in our previous section's derivation.

Another approach is to rewrite the moments of the coordinates as the Hamiltonian operators ${\hat {\vec p}}^2 =\frac{\partial^2}{\partial {t^*}^2}+ {\vec \nabla}^2$, and obtain
\begin{equation}
P(\vec x, \beta)={e^{-\beta [\frac{\partial^2}{\partial {t^*}^2}+ {\vec \nabla}^2 + {\alpha} V (\vec x', \lambda') ]}}
\end{equation}
The partial differential w.r.t. $\frac{\partial}{\partial \beta}$ obtains  for $-\beta \le \lambda \le \beta$
\begin{equation}
\frac{\partial P(\vec x,\lambda)}{\partial \lambda}=\frac{D}{2}\sum\limits_{i=1}^{3} \frac{\partial^2 }{\partial {x_i}} P+\frac{D_t}{2} \frac{\partial^2}{\partial {(-it)} ^2}P + \alpha V(\vec x,\lambda) P(\vec x,\lambda) 
\end{equation}


The Fourier transform $\frac{1}{2\beta}\sum\limits_{0}^{inf} P(\vec x, i\omega_n)\frac{e^{-i \omega_n \lambda}}{2\pi}$ and $\int P(\vec k,i\omega_n)$ transforms the Shroedinger-like PDE when in the two-point form with a delta function source to the Matsubara Green's function
\begin{equation}
P(\vec k,i\omega_n)=\frac{1}{i\omega_n - \epsilon_o (\vec k) - \alpha \Sigma (\vec k ,i\omega_n)}
\end{equation}
The inverse Fourier transform, with the potential self energy equal to zero for simplicity, is the distribution function $P_o (\vec x, \lambda; \vec x', \lambda')=\frac{e^{-\frac{(\vec x - \vec x')^2}{2D\Delta \lambda}}}{\sqrt{4\pi D \Delta \lambda}}$. 

The potential energy can be included via the transformation of coordinates $\vec K^2 = \vec k^2 + \alpha V(\vec k, i\omega_n =0)$ ...the inverse Fourier transform yields a Gaussian $P (\vec X, \hat\lambda; \vec X', \hat\lambda')=\frac{e^{-\frac{(\vec X - \vec X')^2}{2D\Delta \hat\lambda}}}{\sqrt{4\pi D \Delta \hat\lambda}}$, and the coordinate transformation is then to transform $ \vec X -> \vec x $ by direct Fourier transforms, an alternative to PDF transformations, coordinate transformations, and SDE stochastic transformations.

A point of interesting discussion here is of the discrete Fourier transform and the Matsubara frequencies. The sum corresponds to even frequencies or Bosonic Bose-Einstein statistics, and odd frequencies or Fermionic Fermi-Dirac statistics. In this case the $f(\vec k, i\omega_n -> \omega+ i\nu)=\frac{1}{e^{-\beta[\epsilon_o (\vec k)  - \mu]} + 1}$.....The point here is that the momentum vector is a 4Dimensional relativistic vector that includes the real time variable. It is derived from a boson equation that is parametrized with an evolution parameter $\lambda$, here related to the physical temperature by the Matsubara derivation. The generalized 5Dimensional equation of 4D+1p 1parameter yielding a higher dimensional Fermi-dirac statistics and higher dimensional Bose-Einstein statistics is a novel feature that will be investigated further in future work.

The derivation of this curvature potential acting upon Bosonic particles here photons has been instructive in pointing out the connexions to our entropy state function derivation, the stochastic differential derivation, and to this 5D Shroedinger-like relativistic quantum mechanics.

The real curved space-time equation in our derivation would be built from the flat spacetime momentum and therefore wave equation deLambertian, then modified to include the curvature terms as diffusion coefficients obtained from the curved space-time metric. A change of coordinates as before transforms ${ \sum_{l=1}^{3} {B_l (\vec x,\lambda)} {{\nabla}^2}_{l _ {\vec x}}} + { B_{t^*} (\vec x, \lambda)} {\nabla^2}_{t^*} -> { \sum_{l=1}^{3} {{\nabla}^2}_{l _ {\vec X}}} +  {\nabla^2}_{T^*}$ here we write the nabla as a partial derivative for simplicity of notation. The operator form of the momentum is utilized in our maximum entropy derivation, the Matsubara Fokker-Planck or Schroedinger-like 5D or 4D+1p equation obtained, and Fourier transformed to the momentum-frequency space. The subsequent inverse Fourier transformation obtains statistics of Fermions and Bosons and if need be mixed and fractional statistics. We then have side-stepped the Dirac problem of quantizing curved space-time operators. Commutation relations are not considered here as problematic as the transformations of curvature terms which modify flat space-time operators  transforms the problem $entirely$ to an effective linear flat space-time equations with nonlinear potential and diffusion transformed drift potential terms. A new flat-like space-time which is effectively quantizable, effectively solvable in analytic form. Once the quantum mechanics albeit in 5D or 4D+1p spacetime is obtained from the flat spacetime operator representation, the curvature is restored by the analytic forms of the coordinate transformations which it has been discussed in detail can be obtained from the state function, the macroscopic  PDF or the microscopic SDE levels. The remaining problem of interpretation of the parametrization, the evolution parameter $lambda$, and the Lagrange multiplier $\beta$ is the decades old noted correspondence between variance, temperature, diffusion and with properly applied analytic continuation, quantum mechanics. The theory then becomes an entirely well founded physical approach to the quantum statistical mechanics of curved space-time in terms of relativistic thermodynamics. We need not discuss in detail the correspondence between the maximum entropy principle, information theory and thermodynamical variational theory.

\section{Exact coupling of Normal Curved Space-Time and Black-Hole Interior.}
The Hawking problem in obtaining albeit successfully the radiation evaporation dynamics of Black-Holes was to match boundary conditions between the normal curved space-time and the Black-Hole interior. The problem was complicated by a noted boundary conditions matching pathology in elliptical PDEs. A problem or pathology that does not exist in our 5D or 4D+1p Schroedinger-like relativistic mechanics of the parabolic PDE form. The theory of boundary conditions matching has been discussed in detail by T.E. Feuchtwang for the 3D Cartesian and spherical polar and generalized coordinate systems. We adhere to this approach in our $+1$ dimensional theory, and note that the (curvedSpacetime)-(Black-holeInterior) interface heterostructure is a simple two component device in this theory. More complicated strucures would be  (infiniteLead)-(Black-HoleInterior)-(infiniteLead') and here 'lead' is spacetime, these more complicated heterostructures of time bias or space bias driving transport of translation in time-space....note that this more complicated structure is called a worm hole and the time and/or space  bias driving transport of translation in time-space is of the nonequilibrium transport for large scale phenomenon and quantum mechanical  nonequilibrium quantum transport for mesoscopic and nanometer and sub nanometer scales.. At quantum scales these structures are quantum black holes and quantum worm holes, and at Planck scales this more complicated heterostructure is called a GEON after the work of John Archibald Wheeler, for which if a metastable heterostructure of statistical Boson or Fermion or mixed statistics can be analytically derived a theory of fundamental quantum particles can be rigorously obtained.

The simple heterostructure of normal curved spacetime in abrupt coupling to a  Black-Hole is obtained by the delta function coupling method. This is obtained from a Green's theorem derivation as
\begin{eqnarray}
 \sum\limits_{n}^{inf} \int g(\vec x ; \vec x '' ;  i\omega _n)   [i \hbar \omega _n +  {{{\vec \nabla_{\vec x''}}}^2} + H' (\vec x '')]       G(\vec x '' ;\vec x ' ; i \omega _n){d\vec x}''     \\ \nonumber
  - \sum\limits_{n}^{inf} \int G(\vec x ;\vec x '' ; i\omega _n)[i \hbar \omega _n + {{{\vec \nabla_{\vec x''}}}^2} + H'(\vec x'')] g(\vec x '' ;\vec x ' ; i\omega_n) d\vec x'' =0
\end{eqnarray}
These equations can be integrated by parts to obtain the Dirichlet $\\$ $g(\vec x)|_{\vec x =boundary}  = 0$ and the Von Neumann boundary condition Green's functions  $\frac{\partial g(\vec x) }{\partial \vec x} |_{\vec x =boundary}  = 0$. Other mixed boundary conditions are also possible. The derivation is explicitly discussed in \cite{feuchtwang1}. The coupling due to matching boundary conditions can be generalized to delta function coupling Hamiltonian as $h(\vec x)=[\frac{\partial}{\partial \vec x},{\vec\delta}(\vec x- \vec B - \eta) ]_{+} - [\frac{\partial}{\partial \vec x},{\vec\delta}(\vec x + \vec B - \eta) ]_{+}$.

The curved spherical Black-Hole encourages us to work in spherical polar coordinates. We will do that in subsequent derivations. we wish now to point out that the surface Green's functions are immediately obtained, they are for Dirichlet boundary conditions as an example \cite{feuchtwang1}
\begin{equation}
G(r,r',i\omega_n)|_{r=r'=R_{bh}}=\left( \frac{g_{st} (r,r',i\omega_n)g_{bh}(r,r',i\omega_n)}{[g_{st} (r,r',i\omega_n)+g_{bh}(r,r',i\omega_n)]} \right) |_{r=r'=R_{bh}}
\end{equation}

and a similar expression in terms of the 1st order partial derivatives is obtained for the Von Neumann boundary conditions... these surface Green's functions are called the Garcia-Molliner Green's functions. The overall Green's function is written in terms of the normal curved spacetime Green's function $g_{st}$ and the Black-Hole Green's function $g_{bh}$ which both Green's functions exist throughout space-time yet are restricted by the boundary matching we have discussed to the regions of the $r_{st} \ge R_{bh} \ge r_{bh} \ge 0+\eta$

The Green's function is the probability distribution function for analytically continued quantum mechanical functions, and is the probability function for hermitian conjugated functions when complex valued. The Matsubara Green's function is a real valued function as can be noticed from the 5D Schroedinger-like PDE, the solution of which is the Gaussian we have discussed before. the entropy for the Black-Hole is calculated as $<S(\beta)>=\int G(r,r')lnG(r,r')drdr'$. The temperature is set after the inverse Fourier transform as discussed in the previous section, and with choice of Bose-Einstein , Fermi-Dirac or fractional Haldane statistics \cite{haldane1}. The surface entropy is obtained from the surface Green's function as $S(\beta)=GlnG|_{r=r'=R_{bh}}$.

The overall Green's function can be written as 
\begin{equation}
G(r,r',i\omega_n)=g(r,r',i\omega_n)-{\frac{\hbar^2}{2m}}[{g_{st}}(r,r'')\frac{\partial}{\partial r''}G(r'',r')|_{r''={R_{bh}^+}}^{r''={R_{bh}}^-}  \label{eqnblah1}
\end{equation}

The Green's function can be written as a function of the coupling self energy. 
The Green's function with the transpose equation can be evaluated as $\partial^2 _{r'',r'} G|_{}= -\frac{\hbar^2}{2m} \frac{1}{[g_{st} (B,B)+g_{bh} (B,B)]}$. The Green's function can also be written as a function of the coupling self energy, where the self energy is $\Sigma^r = -\frac{\hbar}{2m}(\gamma_{D\alpha} - \gamma_{\alpha})- \frac{1}{g^r _{\alpha} (B,B)}$. The self energy is dependent on the method of partitioning the Hamiltonian, and the Dirichlet boundary conditions reduce this to \cite{fred1} $\Sigma^r _{\alpha}= \frac{\hbar \gamma_{\alpha}}{2m}=\frac{\hbar k_{\alpha}}{2m} $, and where $\alpha = st, bh$ and $k_{\alpha}=\sqrt{\hbar \omega - \Sigma(\vec k_{\alpha} , \omega)  - \Sigma^r _{ \alpha}-i\eta}$, and here we have written it in the analytically continued form $i\omega_n -> \omega + i\eta$.

A possible simplification can be made that focuses on a region of space-time, say the curved normal spacetime, perturbed by coupling to the black hole. This is a transformation that can be derived from the atomistic discrete matrix method, and as we have shown is also a transformation that can be derived from the continuous method. The Green's functions are then the Green's functions derived perturbed by the coupling self-energy such that $G_{st}=\frac{1}{i\hbar\omega_n -\epsilon(\vec k) -\Sigma(\vec k, i\omega_n) -\Sigma_{st} - \Sigma_{bh}}$ and similarly for the Green's function of the black hole.

\section{Nonequilibrium transport}
The equations for the nonequilibrium Green's functions are
\begin{equation}
G^+ = (1+ \Sigma G^r)g^+ (1-\Sigma^a G^a) + G^r \Sigma^+ G^a
\end{equation} 

The current is a characteristic of the nonequilibrium transport. In quantum mesoscopic, nanometer scale interface devices, the device is characterized by the I-V current-voltage. The device current flow is driven by bias, by current, voltage bias, material chemical potential differences. here this is a space-time derivation. What would the driving correspond to? The driving is a mismatch in potential energy between the curved space-time and the black-hole. The analogy to current driving here is a current of quantum particles incident on the blackhole...this can be a current of virtual particles. The analogy to chemical potential is also available for us to utilize in this derivation of the quantum current. This can be seen as the simple inclusion of the number of particles as a non-constant and controlled by the chemical potential as a constant as in $<n>_\alpha=\frac{1}{e^{\beta( \hbar \omega - \mu_\alpha)} \pm   1}$  and as derived from Grand canonical ensemble considerations. Another intriguing analogy is that here these Fermi-Dirac and Bose Einstein statistics are 4D+1p and obtained from the analytical continuation of the Matsubara frequencies summation to the  integration of frequency $\frac{1}{2\beta}\sum\limits_{n=0}^{inf} g(i\omega_n) -> \int f(\omega) g(\omega) \frac{d\omega}{2\pi}$ and $f(\omega)$ is Fermi-Dirac or Bose-Einstein as the even or odd frequency summation \cite{bigbookofmany-particlephysics1}. 
  This can be generalized as by Datta and Maclleland \cite{Maclleland1} , as the nonequilibrium Green's function can be written as $g^+  (\vec k, \omega) =-2if(\hbar \omega- \mu)a(\vec k, \omega)$ for the unperturbed Green's function, and as $G^+ (\vec x, \vec x', \omega) = -2iF(\hbar\omega - \mu(\vec x,\vec x', \hbar \omega) \:\:) A(\vec x, \vec x', \hbar \omega)$. Here the generalized chemical potential is 4vector dependent, and the driving bias due to the time and space curvature can be seen clearly. This is perhaps a more natural representation, as for more complicated structures, say worm holes, the space can be the same (?) however the time can be different, with the bias or chemical potential being different due to the time difference between the two sides of the worm hole.

The current is obtained from the nonequilibrium Green's function as
\begin{eqnarray}
<J(r)>|_{r=B}= \int (\frac{\partial}{\partial r} - \frac{\partial}{\partial r'})G^+ (r,r',\omega)|_{r'->r} \frac{d\omega}{2\pi}\\ \nonumber
=\int \partial_r G^r (r,B)|_{r=B} \Sigma^+ G^a (B,B) - G^r \Sigma^+  \partial_{r'} G^a (B,r')|_{r' = B} \frac{d\omega}{2\pi}
\end{eqnarray}
and here we set the the radial coordinate to the event horizon black hole radius for definiteness. 

We see then that the current is nonzero from the following discussion. The NEGF in the approximation $G^+ = G^r \Sigma^+ G^a$ when differentiated as in the representation Eq.(\ref{eqnblah1}) and its transpose derives the second order partial mixed differential $\partial^2 _{r,r'=B} G^{r,a,+,-}=const/[g_{st} (B,B) + g_{bh} (B,B)]^{r,a,+,-} $, and $\partial_{r'=B} G(B,r') =g(B,B) \partial^2 _{r'',r'} G(r'',r')|_{r'',r' = B}$ and as the Dirichlet boundary conditions rid the uncoupled Green's functions $\partial_{r/r'=B} \:\: g_{\alpha}(r,r')=0$, the nonequilibrium green's function becomes a sum of uncoupled Green's functions only, evaluated at the event horizon radius of the black hole. We make the correspondence that this nonvanishing current evaluated at the radius of the black hole corresponds to the Hawking radiation.

\section{conclusion}
We have derived an exact transport theory for curved spacetime and Black hole heterostructures. This theory is applicable to more complex structures such as worm holes. We have derived a quantum thermal Green's functions representation of curved spacetime and utilized the 4D+1p Schroedinger-like PDEs as the equations of evolution for the rigorous Green's theorem coupling of normal curved spacetime and the Black hole structure interface. We have applied the continuous theory of self energy of coupling to the interface coupling of the spacetime and black hole horizon and obtained the current evaluated at the radius of the black hole. We have shown that even in the approximation of $G^+ =G^r \Sigma^+ G^a$ corresponding to no large nonequilibrium term contributions the current at the black hole horizon is nonzero. We make the correspondence that this nonvanishing current corresponds to the Hawking radiation. In the future we will derive the full nonequilibrium contribution, thereby the change in rate of evaporation of the black hole. Also, numerical simulations will help visualize the dynamics of this theory. also of interest and for future work is the application of this theory to the worm hole physics, and the exploration of the limitations if any of the quantum length scales of application of this approach to highly curved spacetime.

\pagebreak

\end{document}